\documentclass[prl,twocolumn,aps]{revtex4}
\usepackage{graphicx}
\font\mybbb=msbm10
\def\Bbb#1{\mbox{\mybbb #1}}

\begin{document}

\title[Entanglement For $XY$ Model]{Entanglement  in $XY$ Spin Chain}

\author{A.\ R.\ Its}
\affiliation{Department of Mathematical Sciences, Indiana
University-Purdue University Indianapolis, Indianapolis, IN
46202-3216}

\author{B.-Q. Jin}
\author{V.\ E.\ Korepin}
\affiliation{C.N.\ Yang Institute for Theoretical Physics, State
 University of New York at Stony Brook, Stony
Brook, NY 11794-3840}
\date{\today}

\begin{abstract}
We consider the ground state of the  $XY$ model on an infinite chain at zero
temperature.
Following  Bennett,  Bernstein,  Popescu, and Schumacher we use entropy of a
sub-system as  a measure of entanglement. Vidal, Latorre,
Rico and  Kitaev conjectured that  von Neumann entropy of a large block of
neighboring spins approaches a constant as the size of the block increases.
We evaluated  this limiting  entropy  as a function of anisotropy and
 transverse magnetic field.
We used the methods based on  integrable Fredholm 
operators and Riemann-Hilbert problem. The entropy is singular at phase
 transitions.
\end{abstract}


\maketitle

There is  an essential interest in quantifying entanglement
in various quantum systems 
\cite{hlw,zanardi,fazio,fan,rasetti,honk, nielsen,V,vidal, julien,salerno,LRV,jin,K,cardy, kais,eisert,ABV,VMC,LO,PP,FS,V2}.
Entanglement is a fundamental measure of 'quantumness' of the system:  
how much quantum effects we can observe and use. It is the primary
 resource in
quantum computation and quantum information processing \cite{BD,L}.
Stable, large scale entanglement  is necessary for 
scalability of  quantum  computation \cite{rasetti, zanardi}.
For  experimental demonstration one can look, for example in  \cite{GRAC,V}.
The  $XY$ model in a transverse magnetic was studied from the point for view of
 quantum information in \cite{fazio,vidal}, \cite{keat},
 \cite{yang,sun}. 
It was  conjectured in  \cite{vidal} that in $XY$ and other gapped models
 the entropy of a block of  $L$  neighboring spins approaches a constant
as  $L\rightarrow \infty $ (see comment \footnote{For  AKLT-VBS models this
 was proved in \cite{fan}}).
In this paper we  evaluated the entropy of a  block of $L$  neighboring 
spins in the ground state of $XY$ model in the limit $L\rightarrow \infty $ 
analytically, see  (\ref{33}).
Hamiltonian of $XY$  model can be written as
\begin{eqnarray}
H=-\sum_{n=-\infty}^{\infty}
(1+\gamma)\sigma^x_{n}\sigma^x_{n+1}+(1-\gamma)\sigma^y_{n}\sigma^y_{n+1}
+ h\sigma^z_{n} \label{xxh}
\end{eqnarray}
Here $\gamma$ is  anisotropy parameter ($0<\gamma<1$) ;  $\sigma^x_n$, $\sigma^y_n$  $\sigma^z_n$ are
Pauli matrices  and  $h$ is a  magnetic field.
The  model was solved  in \cite{Lieb},  \cite{mccoy}, \cite{mccoy2}, \cite{gallavotti}.
Toeplitz determinants were used for  evaluation of some correlation functions,
see  \cite{tak} and  \cite{aban}.
Integrable Fredholm operators  were used for calculation of other correlations,
 see   \cite{sla, dz, izer, pron}.

The ground state of the model $|GS\rangle$ is 
unique. So the entropy of the whole infinite ground state is zero,
but it can be positive for a subsystem [a part of the ground state].
We shall calculate the entropy of a block of  $\mathrm{L}$ neighboring spins.
We can think that  the ground state is a binary system  
$|GS\rangle = |A \& B\rangle $.
We can call the block of  $L$ neighboring spins by  sub-system A
and the rest of the ground state by  sub-system B.
The  density matrix of the ground state is $|GS\rangle \langle GS|$.
We shall denote it by \mbox{$\rho_{AB}=|GS\rangle \langle GS|$}.
The density matrix of the $L$ neighboring spins [subsystem A] is
 \mbox{$\rho_A=
Tr_B(\rho_{AB})$}.
Von Neumann entropy $S(\rho_A)$  of the subsystem A
can be represented as following:
\begin{eqnarray}
S(\rho_A)=-Tr_A(\rho_A \ln \rho_A), \label{edif}
\label{olds}
\end{eqnarray}
This entropy defines the dimension of the  Hilbert space of states of the
 block of $L$ spins.
Majorana operators were used in \cite{vidal} to describe the entropy (\ref{olds})  by the following matrix:
\begin{eqnarray}
\mathbf{B}_L=\left( \begin{array}{cccc}
\Pi_0 &\Pi_{-1}& \ldots &\Pi_{1-L}\\
\Pi_{1}& \Pi_0&   &   \vdots\\
\vdots &      & \ddots&\vdots\\
\Pi_{L-1}& \ldots& \ldots& \Pi_0
\end{array}     \right) \nonumber
\end{eqnarray}
Here
$$
\Pi_l=\frac{1}{2\pi} \int_{0}^{2\pi} \, \mathrm{d} \theta\,
e^{-\mathrm{i} l \theta} {\cal G}(\theta),\quad {\cal
G}(\theta)=\left( \begin{array}{cc}
               0& g(\theta)\\
               -g^{-1}(\theta)&0
               \end{array} \right)
               $$
\begin{equation}\textrm{and} \qquad g(\theta)=\frac{\cos \theta -\mathrm{i}
\gamma\sin \theta -h/2}{|\cos \theta -\mathrm{i} \gamma\sin \theta
 -h/2|} \quad . \end{equation}

\begin{figure}[ht]
\begin{center}
\includegraphics[width=3in,clip]{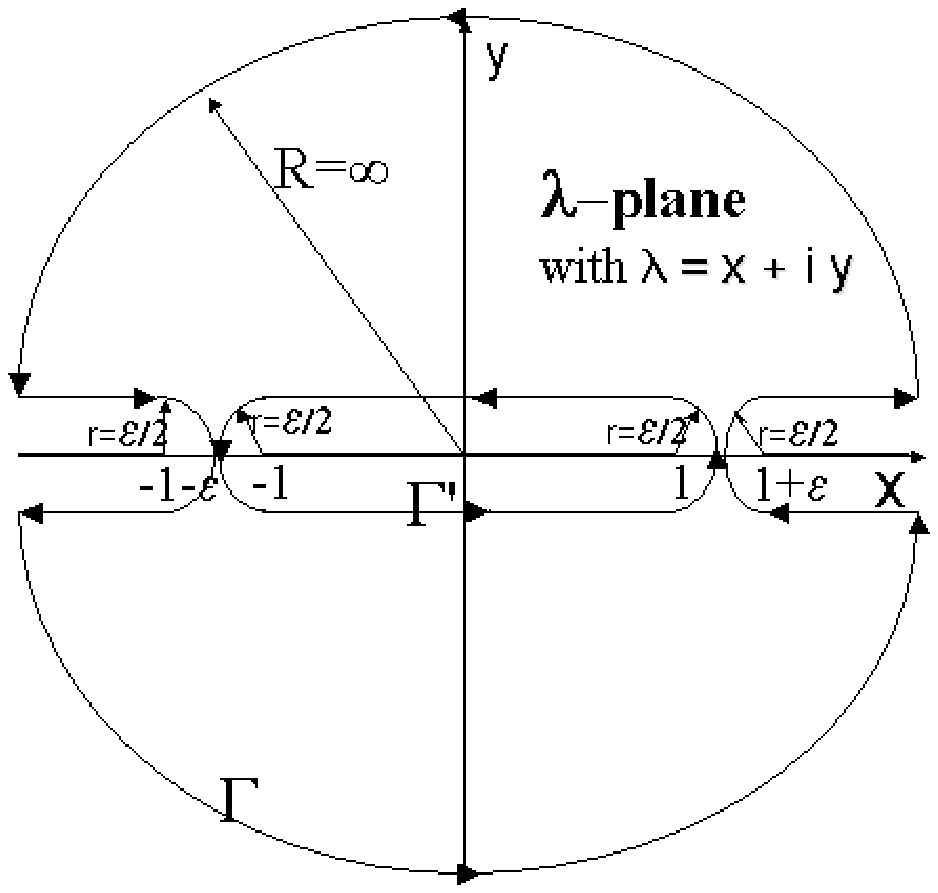}
\end{center}
\caption{\it Contours \mbox{$\Gamma'$} (smaller one)
and \mbox{$\Gamma $} (larger one). Bold lines $(-\infty,
-1-\epsilon)$ and $(1+\epsilon,\infty)$ are the cuts of integrand
$e(1+\epsilon,\lambda)$. Zeros of $D_{L}(\lambda)$
(Eq.~$\ref{exd}$) are located on bold line $(-1, 1)$. The arrow is
the direction of the route of integral we take and $\mathrm{r}$
and $\mathrm{R}$ are the radius of circles. $\P $  } \label{fig1}
\end{figure}

One can use an orthogonal matrix  $V$ to transform
$\mathbf{B}_L$ to a canonical form:
\begin{eqnarray}
V \mathbf{B}_{L} V^T= \oplus_{m=1}^{L} \nu_m \left(
\begin{array}{cc}
               0& 1\\
               -1&0
               \end{array} \right),\label{vd}
               \end{eqnarray}
               The real numbers  $-1<\nu_m<1$  play an important role.
               We shall call them eigenvalues.
               The entropy of a block of  $L$ neighboring
               spins was represented  in \cite{vidal} as
\begin{eqnarray}
S(\rho_A)&=&\sum_{m=1}^{L} H(\nu_m) \label{eaap1}
\end{eqnarray}
with
\begin{eqnarray}
 H( \nu)= -\frac{1+\nu}{2} \ln \frac{1+\nu}{2}-\frac{1-\nu}{2} \ln \frac{1-\nu}{2}.\label{intee1}
\end{eqnarray}
In order to calculate the asymptotic form of the entropy  let us  introduce:
\begin{eqnarray}
\widetilde{\mathbf{B}}_{L}(\lambda)=\mathrm{i}\lambda I_{L}-
\mathbf{B}_{L}, \quad D_{L}(\lambda)=\det
\widetilde{\mathbf{B}}_{L}(\lambda)
\end{eqnarray}
and \begin{eqnarray}
 e(x, \nu)= -\frac{x+\nu}{2} \ln \frac{x+\nu}{2}-\frac{x-\nu}{2} \ln \frac{x-\nu}{2}.\label{intee}
\end{eqnarray}
 Here $I_{L}$ is the
identity matrix of dimension $2L$. By definition, we have
$H(\nu)=e(1,\nu)$ and
\begin{eqnarray}
D_{L}(\lambda)=(-1)^{L} \prod_{m=1}^{L} (\lambda^2-\nu_m^2).
\label{exd}
\end{eqnarray}
In \cite{jin} we used  Cauchy residue theorem  to rewrite
formula ($\ref{eaap1}$)  in the following form:
\begin{eqnarray}
S(\rho_A)=\lim_{\epsilon \to 0^+} \frac{1}{4\pi \mathrm{i}}
\oint_{\Gamma'} \mathrm{d} \lambda\,  e(1+\epsilon, \lambda)
\frac{\mathrm{d}}{\mathrm{d} \lambda} \ln
D_{L}(\lambda)\;.\label{eaa}
\end{eqnarray}
Here the contour \mbox{$\Gamma'$} in Fig~$\ref{fig1}$ encircles
all zeros of \mbox{$D_{L}(\lambda)$}.
\noindent We also realized that
$\widetilde{\mathbf{B}}_{L}(\lambda)$ is a block Toeplitz matrix
with the generator $\Phi(z)$, i.e.
\begin{eqnarray}
\widetilde{\mathbf{B}}_L(\lambda)=\left( \begin{array}{cccc}
\widetilde{\Pi}_0 &\widetilde{\Pi}_{-1}& \ldots &\widetilde{\Pi}_{1-L}\\
\widetilde{\Pi}_{1}& \widetilde{\Pi}_0&   &   \vdots\\
\vdots &      & \ddots&\vdots\\
\widetilde{\Pi}_{L-1}& \ldots& \ldots& \widetilde{\Pi}_0
\end{array}     \right) \quad \textrm{with}\nonumber
\end{eqnarray}
\begin{equation}
 \widetilde{\Pi}_l=\frac{1}{2\pi\mathrm{i}}\oint_{\Xi} \, 
\mathrm{d} z\, z^{-l-1} \Phi(z), \quad \Phi(z)=\left( \begin{array}{cc}
               \mathrm{i}\lambda & \phi(z)\\
               -\phi^{-1}(z)&\mathrm{i}\lambda
               \end{array} \right) \label{defphi}
\end{equation}
\begin{equation}
\textrm{and}\quad \phi(z)=
\left(\frac{\lambda_1^*}{\lambda_{1}}\frac{(1-\lambda_1\,
z)(1-\lambda_2\, z^{-1})}{(1-\lambda_1^* \,
z^{-1})(1-\lambda_2^*\, z)}\right)^{1/2}
\end{equation}

\begin{figure}[ht]
\begin{center}
\includegraphics[width=3in,clip]{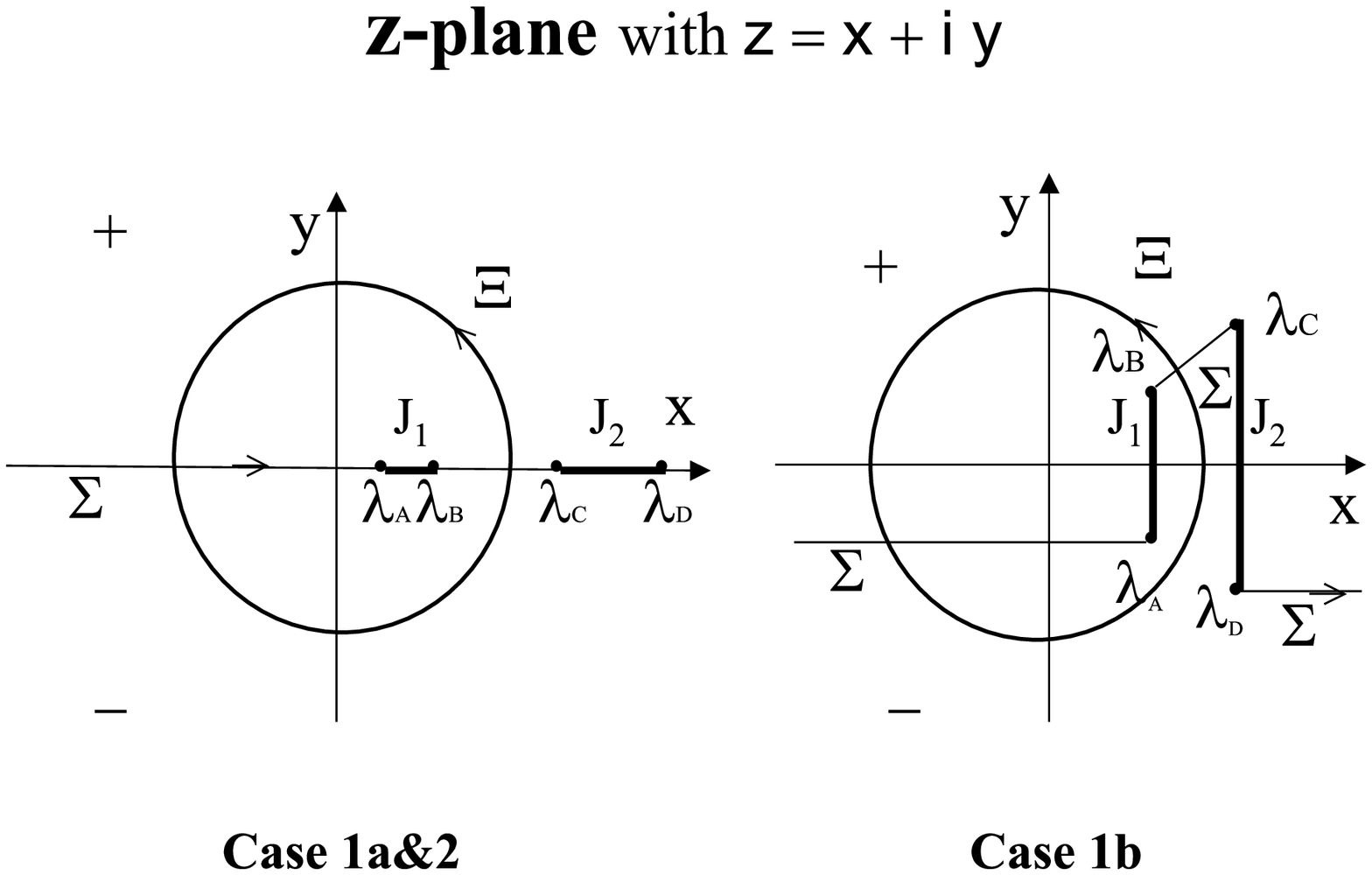}
\end{center}
\caption{\it Polygonal line $\Sigma$ (direction as labeled) separates
the complex $z$ plane into the two parts: the part $\Omega_{+}$
which lies to the left of $\Sigma$, and the part  $\Omega_{-}$
which lies to the right of $\Sigma$. Curve $\Xi$ is the unit circle in
anti-clockwise
direction. Cuts $J_1, J_2$ for functions $\phi(z),w(z)$ are
labeled by bold on line $\Sigma$. Definition of the end points of the cuts 
$\lambda_{\ldots}$ depends on the case:
{\bf Case} $1$a:  $\lambda_A=\lambda_1$ and
$\lambda_B=\lambda_2^{-1}$, $\lambda_C= \lambda_2$ and $\lambda_D=
\lambda_1^{-1}$. {\bf Case} $1$b:
 $\lambda_A=\lambda_1$ and
$\lambda_B=\lambda_2^{-1}$, $\lambda_C= \lambda_1^{-1}$ and
$\lambda_D= \lambda_2$. {\bf Case } $2$:  
 $\lambda_A=\lambda_1$ and $\lambda_B=\lambda_2$, $\lambda_C=
\lambda_2^{-1}$ and $\lambda_D= \lambda_1^{-1}$. $\P$} \label{fig2}
\end{figure}
We fix the branch by requiring that $ \phi(\infty)>0$.
We use $*$ to denote complex conjugation and $\Xi$ is the unite circle shown in
Fig.~\ref{fig2}. The points $\lambda_1$, $\lambda_2$ are different
depending on the {\bf Case}:
\begin{enumerate}
\item{ {\bf Case} $1$a: $\quad 2\sqrt{1-\gamma^2}<h< 2$}.
\item{{\bf Case} $1$b:  $\quad 0<h<2\sqrt{1-\gamma^2}$ }.
\item{{\bf Case } $2$: $\quad h> 2$  }.
\end{enumerate}

\noindent   In two cases $1$a and $2$:
\begin{eqnarray}
\lambda_1=\frac{h-\sqrt{h^2-4(1-\gamma^2)}}{2(1+\gamma)},\quad
\lambda_2=\frac{1+\gamma}{1-\gamma} \lambda_1.\label{ldef1}
\end{eqnarray}
These $\lambda$ are real because  $h^2>4(1-\gamma^2)$ for both cases.

\noindent  In case $1$b:
\begin{eqnarray}
\lambda_1=\frac{h-\mathrm{i} \sqrt{4(1-\gamma^2)-h^2}}{2
(1+\gamma)},\quad
 \lambda_2=1/\lambda_1^*.\label{ldef2}
\end{eqnarray}
In this case $h^2<4(1-\gamma^2)$.

(Note that in the Case $1$ the poles of the function $\phi(z)$ coincide
with the points $\lambda_{A}$ and $\lambda_{B}$, while in the Case $2$
they coincide with the points $\lambda_{A}$ and $\lambda_{C}$.)

By virtue of Eq.~(\ref{eaa}), our objective becomes the asymptotic
 calculation of the determinant of block Toeplitz matrix
$D_L(\lambda)$  or, rather,
its $\lambda$ -derivative $\frac{d}{d\lambda}\ln D_L(\lambda)$. A general
asymptotic representation of the determinant of a block
Toeplitz matrix, which generalizes the classical strong Szeg\"o theorem to
the block matrix case, was obtained by Widom in \cite{widom} (see also
more recent work \cite{bottcher} and references therein). The important
 difference with the scalar case is the non-commutativity of the associated
Weiner-Hopf factorization. This creates serious technical problems. In our
 work we circumvent this obstacle by
using an alternative approach to Toeplitz determinants
suggested by Deift in \cite{deift}. It is based
on the Riemann-Hilbert technique of the theory of
``integrable integral operators'', which was
developed in \cite{iiks}, \cite{korepin} for evaluation of correlation functions of quantum
completely integrable [exactly solvable] models
(see also comment {\footnote{In its turn, the approach of \cite{iiks} is based on
the ideas of  \cite{jmms}. Several principal aspects of the
integrable operator theory,
especially the ones concerning with the integrable differential systems
appearing in random matrix theory,
have been developed in \cite{tw}. Some of the important elements of  modern theory of
integrable operators were already implicitly present in the
earlier work \cite{sakh}.}}).
It turns out that, using the block matrix
version of \cite{iiks} suggested in \cite{hi}, one can
generalize Deift's scheme to the block Toeplitz matrices.
In addition, we were able to find the explicit Weiner-Hopf
factorization of the generator $\Phi(z)$ which eventually made it
possible to perform an explicit evaluation of the asymptotic of the entropy
$S(\rho_{A})$. The final result is given in terms of
elliptic functions and is presented in  Eq.~(\ref{333}) below.
In what follows we shall outline our calculation providing
the necessary facts concerning integrable Fredholm operators.
More details,
including the evaluation of error terms, will be presented
in a separate publication.

Let $f_{j}(z)$ and $h_{j}(z)$, $j = 1, 2$, be $2\times 2$ matrix
functions. We introduce the class of
{\bf integrable operators} $K$  defined on
$L_{2}(\Xi, {\Bbb C}^2)$ by the following equations (cf. \cite{hi}),
\begin{equation}
(K\, X)(z) = \oint_{\Xi}K(z,z')\, X(z')dz'\quad
\textrm{for}\quad X\in L_{2},
\end{equation}
where
$$
K(z,z') =
\frac{f^{T}(z)h(z')}{z-z'},\quad f(z) = \left(
\begin{array}{c}
f_{1}^{T}(z)\\
f_{2}^{T}(z)
\end{array}\right)
$$
\begin{equation}\label{intkernel}
h(z) = \left( \begin{array}{c}
h_{1}(z)\\
h_{2}(z) \end{array}\right).
\end{equation}
Let $I_{2}$ denote the $2\times 2$ identity matrix. Put
\begin{equation}
f_{1}(z) = z^L I_{2}, \quad f_{2}(z) = I_{2} \label{f12}
\end{equation}
\begin{equation}\label{h12}
h_{1}(z) = z^{-L}\frac{I_{2} - \Phi(z)}{2\pi i},\quad h_{2}(z) =
-\frac{I_{2} - \Phi(z)}{2\pi i}.
\end{equation}
Then, essentially repeating the arguments of \cite{deift}, we have
the following relation
\begin{equation}\label{fredholm}
D_L(\lambda) = \det (I - K),
\end{equation}
So we represented $D_L(\lambda) $ as a Fredholm determinant  of
the integral operator $K$.
Define the resolvent operator $R$ by $$(I-K)
(I+R)=I.$$ Here
$I$ is the identity operator in $L_{2}(\Xi, {\Bbb C}^2)$. Then we have
the general equation,
$$
\frac{d}{d\lambda}\ln D_L(\lambda) = -\mbox{Tr}\,\left[
(I-K)^{-1}\frac{d}{d\lambda}K\right],
$$
which, taking into account that in our case
$$
\frac{d}{d\lambda}K(z, z') = -iK(z,z')(I_{2} - \Phi(z'))^{-1},
$$
can be rewritten as
\begin{equation}\label{trace1}
\frac{d}{d\lambda}\ln D_L(\lambda)
= i\oint_{\Xi}\mbox{tr}\,\left[R(z,z)(I_{2} -
\Phi(z))^{-1}\right]\, dz.
\end{equation}
In the formulae above, ``Tr'' means the trace
taking in the space $L_{2}(\Xi, {\Bbb
C}^2)$, while ``tr'' is the $2\times 2$ matrix trace.
An important general fact is that the resolvent kernel
satisfies the equation (see e.g. \cite{hi}),
\begin{equation}\label{RFH}
R(z,z) = \frac{dF^{T}(z)}{dz}H(z).
\end{equation}
In this equation, the $4\times 2$  matrix functions $F(z)$ and
$H(z)$ are determined by the relations,
\begin{equation}\label{RHF}
F(z) = Y_{+}(z)f(z), \quad z\in \Xi,
\end{equation}
\begin{equation}\label{RHH}
H(z) = (Y^{T}_{+})^{-1}(z)h(z), \quad z\in \Xi,
\end{equation}
where the $4\times 4$ matrix function $Y_+(z)$  can be found
from the (unique) solution of the following {\bf
Riemann-Hilbert problem}:
\begin{enumerate}
\item $Y(z)$ is analytic for $z\notin \Xi$. \item $Y(\infty) =
I_{4}$, where $I_{4}$ denotes the $4\times 4$ identity matrix.
\item $Y_{-}(z) = Y_{+}(z)J(z)$ for $z\in \Xi$ where $Y_{+}(z)$
($Y_{-}(z)$) denotes the left (right) boundary value of $Y(z)$ on
unit circle $\Xi$ (Note: ``+'' means from inside of the unit
circle). The $4\times 4$ jump matrix $J(z)$ is defined by the
equations,
$$
J(z) = I_{4} + 2\pi if(z)h^{T}(z)
$$
\begin{equation}\label{J} = \left( \begin{array}{cc}
2I_{2} - \Phi^{T}(z)  & -z^{L}(I_{2} - \Phi^{T}(z))\\
z^{-L}(I_{2} - \Phi^{T}(z))& \Phi^{T}(z)\end{array}\right).
\end{equation}
\end{enumerate}

Eqs.~(\ref{trace1}) and (\ref{RFH}) reduce the original
question to the asymptotic analysis of the solution $Y(z)$ of
the Riemann-Hilbert problem (1-3). Our observation is that
once again we can generalize the  arguments of \cite{deift}
to the case of matrix generator $\Phi(z)$ and arrive to the
following asymptotic solution of the
problem (1-3) ( $L \to \infty$):
\begin{equation}\label{asymp2}
Y_{+}(z)=\left( \begin{array}{cc}
U_{+}^{T}(z)   & ~~-z^{L}U_{+}^{T}(z)\, M(z)\\
0_{2}& (V_{+}^{T})^{-1}(z)\end{array}\right)
\end{equation}
and
\begin{equation}\label{asymp3}
(Y_{+})^{-1}(z)= \left( \begin{array}{cc} (U_{+}^{T})^{-1}(z) & ~~
z^{L}  M(z)\, V_{+}^{T}(z)\\
0_2& V_{+}^{T}(z)\end{array}\right).
\end{equation}
Here $$M(z)=I_{2} - (\Phi^{T})^{-1}(z)$$ and $U_{\pm}(z)$ and
$V_{\pm}(z)$ are  $2\times 2$ matrices solving
the  Weiner-Hopf factorization problem :
\begin{description}
\item {(i)}\quad $\Phi(z)=U_+(z)U_-(z)=V_-(z)V_+(z) \label{wiener-hopf} ,
\quad z\in \Xi $ \item {(ii)} \quad $U_-(z)$ and $V_-(z)$ ($U_+(z)$ and
$V_+(z)$) are analytic outside (inside) the unit circle $\Xi$.
\item {(iii)} \quad $U_-(\infty)=V_-(\infty)=I$.
\end{description}

We can use Eqs.~(\ref{asymp2}) and (\ref{asymp3}) in
Eqs.~(\ref{trace1}) - (\ref{RHH}) and obtain the following
asymptotic formula:
\begin{eqnarray}\label{asymp}
\frac{d}{d\lambda}\ln  D_L(\lambda) =
-\frac{2\lambda}{1-\lambda^{2}}L +\frac{1}{2\pi}
\int_{\Xi}\mathrm{tr}\, \left[\Psi(z)\right]\, dz,\label{dln}
\end{eqnarray}
\begin{equation}
\Psi(z)=\left[ U_{+}'(z)U_{+}^{-1}(z)
+V_{+}^{-1}(z)V_{+}'(z)\right] \Phi^{-1}(z),
\end{equation}
as $L \to \infty$ (see also comment
{\footnote{This formula plays in our analysis the role of
the strong Szeg\"o theorem. We think it would be of interest
to understand its meaning in context of the general
result of Widom \cite{widom}.}}). Here $'$ means a derivative in
 $z$ variable.

By explicit calculation, one can find that
\begin{eqnarray}
(1-\lambda^2) \sigma_3 \Phi^{-1}(z) \sigma_3= \Phi(z),\quad
\sigma_3= \left( \begin{array}{cc} 1 &
0\\
0& -1\end{array}\right).
\end{eqnarray}
Hence,
\begin{eqnarray}
V_-(z)&=&\sigma_3 U_-^{-1}(z)\sigma_3\\
V_+(z)&=&\sigma_3 U_+^{-1}(z)\sigma_3(1-\lambda^2),\quad
\lambda\neq \pm 1,
\end{eqnarray}
and one only needs the explicit expressions for $U_{\pm}(z)$.

Our last principal observation is that, for all $\lambda$
outside of a certain discrete subset of the interval
$[-1, 1]$, the solution to the auxiliary Riemann-Hilbert problem
(i-iii) exists; moreover, the functions
$U_{\pm}(z)$ can be expressed in terms of the Jacobi
theta-functions. Indeed, the auxiliary Riemann-Hilbert
problem (i-iii) can be easily reduced to a type of the
``finite-gap'' Riemann-Hilbert problems which
have already appeared in the analysis of the
integrable statistical mechanics models
(see \cite{diz}). Before we give detail
expressions, let us first define some basic objects:
\begin{eqnarray}
w(z)&=&\sqrt{(z-\lambda_1)
(z-\lambda_2)(z-\lambda_2^{-1})(z-\lambda_1^{-1})},\label{notion1}\\
\beta(\lambda)&=&\frac{1}{2\pi i}\ln
\frac{\lambda+1}{\lambda-1},\end{eqnarray} where $w(z)$ is
analytic on the domain ${\Bbb C} \backslash \, \{ J_1\cup J_2\}$
shown in Fig.~\ref{fig2} and fixed by the condition:
$w(z) \to z^2$ as

 $z \to \infty $.
Next we define
\begin{eqnarray}
\tau=\frac{2}{c} \int_{\lambda_B}^{\lambda_C}\frac{\mathrm{d}
z}{w(z)}, \quad
c=2\int_{\lambda_A}^{\lambda_B}\frac{\mathrm{d}z}{w(z)},\label{important}\end{eqnarray}
\begin{eqnarray}
\delta=\frac{2}{c} \left(-\pi i-\int_{\lambda_A}^{\lambda_B}
\frac{z \mathrm{d} z}{w(z) }\right),  \quad \omega(z)= \frac{1}{c}
\int_{\lambda_A}^{z}\frac{\mathrm{d} z}{w(z)},
\end{eqnarray}
\begin{equation}
 \Delta(z)=\frac{1}{2}\int_{\lambda_A}^z
\frac{z+\delta}{w(z)}\mathrm{d} z,\quad
\kappa=\int_{\lambda_A}^{\infty}\mathrm{d} \omega(z),
\end{equation}
Points $\lambda_A,\lambda_B,\lambda_C, \lambda_D$
and cuts $J_1$, $J_2$ and curves $\Sigma$ and $\Xi$ are shown in
Fig.~\ref{fig2}. We shall also need:
\begin{equation}
\Delta_0=\lim_{z\to \infty}\left[\Delta(z)-\frac{1}{2} \ln
(z-\lambda_1)\right].\label{notion2}
\end{equation}
Here, the contours of integration
for $c$ and $\delta$ are taken along the
left side of the cut $J_{1}$. The contour of
integration for $\tau$ is the segment
$[\lambda_{B}, \lambda_{C}]$. The contours of integration
for $\kappa$ and in (\ref{notion2}) are taken along the line $\Sigma$ to the left
from $\lambda_{A}$; also in (\ref{notion2}), $\arg(z-\lambda_{1}) = \pi$. 
The contours of integration in the
integrals
$\Delta(z)$ and $\omega(z)$ are
taking according to the rule: The contour lies entirely in
the domain $\Omega_{+}$ ($\Omega_{-}$) for $z$ belonging to $\Omega_{+}$
($\Omega_{-}$). It also worth noticing that
$i\tau < 0$.

Now we are ready to introduce the Jacobi theta-function,
\begin{eqnarray}
\theta_3(s)=\sum_{n=-\infty}^{\infty} e^{\pi i \tau n^2+2\pi i s
n}.\label{jac}
\end{eqnarray}
We remind the following properties of this theta-function (see e.g. \cite{ww}):
\begin{eqnarray}
\theta_3(-s)=\theta_3(s),\quad \theta_3(s+1)=\theta_3(s) \label{theta1}\\
\theta_3(s+\tau)=e^{-\pi i \tau-2\pi i s} \theta_3(s) \label{theta2}\\
\theta_3\left(n+m\tau+\frac{1}{2}+\frac{\tau}{2}\right)=0,\quad
n,m\in {\Bbb Z} \label{thetazeros}
\end{eqnarray}
We also introduce the $2\times 2$
matrix valued function $\Theta(z)$ with the entries,
\begin{eqnarray}&&\Theta_{11}(z)=(z-\lambda_1)^{-\frac{1}{2}} e^{\Delta(z)}\nonumber\\&&\quad \quad \quad  \times \frac{\theta_3\left(\omega(z)+\beta(\lambda)-\kappa +\frac{\sigma \tau}{2}\right)}{\theta_3\left(\omega(z) + \frac{\sigma \tau}{2}\right)}\nonumber\\ &&\Theta_{12}(z)=-(z-\lambda_1)^{-\frac{1}{2}} e^{-\Delta(z)}\nonumber\\ &&\quad \quad \quad \times \frac{\theta_3\left(\omega(z)-\beta(\lambda)+\kappa -\frac{\sigma \tau}{2}\right)}{\theta_3\left(\omega(z) - \frac{\sigma \tau}{2}\right)}\nonumber\\&&\Theta_{21}(z)=-(z-\lambda_1)^{-\frac{1}{2}} e^{-\Delta(z)}\nonumber\\ &&\quad \quad \quad \times \frac{\theta_3\left(\omega(z)+\beta(\lambda) +\kappa -\frac{\sigma \tau}{2}\right)}
{\theta_3\left(\omega(z) - \frac{\sigma \tau}{2}\right)}\nonumber\\ &&\Theta_{22}(z)=(z-\lambda_1)^{-\frac{1}{2}} e^{\Delta(z)}\nonumber\\ &&\quad \quad \quad \times \frac{\theta_3\left(\omega(z)-\beta(\lambda)-\kappa +\frac{\sigma \tau}{2}\right)}{\theta_3\left(\omega(z) + \frac{\sigma \tau}{2}\right)},\label{thetad}
\end{eqnarray}
where  $\sigma = 1$ in  Case $1$ and $\sigma =0$ in Case 2,
and $\beta(\lambda)$, $\omega(z)$ and $\kappa$ are  defined in
Eqs.~(\ref{notion1}-\ref{notion2}).   The branch of
$(z-\lambda_1)^{-\frac{1}{2}}$ is defined on the
$z$-plane cut along the part of the line $\Sigma$ which
is to the right of $\lambda_{1}\equiv \lambda_{A}$, and it is fixed by the
condition
$\arg (z-\lambda_1) = \pi, \quad \mbox{if}\quad  z - \lambda_{1} < 0$.

The matrix function $\Theta(z)$ is defined on ${\Bbb C} \backslash
\, {\Sigma}$. However, analyzing the jumps of
the integrals $\omega(z)$ and $\Delta(z)$ over
the line $\Sigma$ and taking into account the
properties (\ref{theta1}) and (\ref{theta2})
of the theta function, one can see that $\Theta(z)$
is actually extended to the analytic
function defined on ${\Bbb C} \backslash \, \{ J_1\cup J_2\}$.
Moreover, it satisfies the jump relations
\begin{eqnarray}
\Theta_+(z)=\Theta_-(z) \sigma_1\qquad z\in J_1\\
\Theta_+(z)=\Theta_-(z)\Lambda \sigma_1\Lambda^{-1}\quad
z\in J_2.\\
\Lambda=i\left(\begin{array}{cc}
              \lambda+1&0\\
              0&\lambda-1
              \end{array}\right),\quad \sigma_1=\left(\begin{array}{cc}
              0&1\\
              1&0
              \end{array}\right)
\end{eqnarray}
Also note:
\begin{eqnarray}
\Theta_{11}(\infty)&=&e^{\Delta_0}\frac{\theta_3\left(\beta(\lambda)+
\frac{\sigma \tau}{2}\right)}
{\theta_3\left(\kappa + \frac{\sigma \tau}{2}\right)}\\
\Theta_{22}(\infty)&=&e^{\Delta_0}\frac{\theta_3\left(\beta(\lambda) -
\frac{\sigma \tau}{2}\right)}
{\theta_3\left(\kappa + \frac{\sigma \tau}{2}\right)}\\
\Theta_{12}(\infty)&=&\Theta_{21}(\infty)=0\,\,,
\end{eqnarray}
and
\begin{equation}\label{detTheta}
\det \Theta(z) \equiv \phi(z) \det
\Theta(\infty)\sqrt{\frac{\lambda_{2}}{\lambda_{1}}}.
\end{equation}
The latter equation follows from the comparison of the jumps
and singularities of its  sides.
Finally, we introduce the matrix
\begin{eqnarray}
Q(z)=\left( \begin{array}{cc}
\phi(z) & -\phi(z)\\
i&i
\end{array}     \right)\label{td}
\end{eqnarray}
Note that $Q(z)$ diagonalizes original jump matrix $\Phi(z)$:
\begin{equation}\label{factoriz}
\Phi(z)=Q(z)\Lambda Q^{-1}(z)
\end{equation}
and $Q(z)$ is analytic on ${\Bbb C}\backslash\, \{J_1\cup J_2\}$
and
\begin{equation}
Q_+(z)=Q_-(z)\sigma_1, \quad z\in J_1\cup J_2.
\end{equation}
We are now ready to present  the solution $U_{\pm}(z)$
of the Riemann-Hilbert problem (i-iii).
Put \begin{equation}A=Q(\infty)
\Lambda^{-1}\Theta^{-1}(\infty).\end{equation} Then,
\begin{eqnarray} U_-(z)=A \Theta(z) \Lambda Q^{-1}(z),\quad |z|\ge 1 \label{u-d1}\\
U_+(z)=Q(z)\Theta^{-1}(z) A^{-1},\quad |z|\le 1.\label{u-d}
\end{eqnarray}
Indeed, by virtue of Eq.~(\ref{factoriz}), we only need
to be sure that $U_-(z)$ and $U_+(z)$ are
analytic for $|z|>1$ and $|z|<1$ respectively. From the
jump properties of $\Theta(z)$ and $Q(z)$ it follows that
$U_{\pm}$ have no jumps across $J_{1,2}$, and hence they might
have only possible isolated singularities at $\lambda_{1,2},\lambda_{1,2}^{-1}$.
The analyticity at these points can be shown by observing that the
singularities, which the functions $\Theta(z)$ and
$Q(z)$ do have at the end points of the segments
$J_{1,2}$, are canceled out in the products (\ref{u-d1})-(\ref{u-d}).

The excluded values of $\lambda$ for which the above construction fails
are $\lambda = \pm 1$ and, in view of Eq.~(\ref{detTheta}), the zeros of
$\theta_3\left(\beta(\lambda)+\frac{\sigma \tau}{2}\right)$, i.e. (see
(\ref{thetazeros})), 
\begin{equation}\label{zeros}
\pm \lambda_{m}, \quad \lambda_{m} =
\tanh \left(m + \frac{1-\sigma}{2}\right)\pi \tau_{0}, \quad m \geq 0,
\end{equation}
where,
$$
\tau_{0} = -i\tau = -i 
\frac{\int_{\lambda_{B}}^{\lambda_{C}}\frac{dz}{w(z)}}
{\int_{\lambda_{A}}^{\lambda_{B}}\frac{dz}{w(z)}}>0 .
$$
Using explicit formulae (\ref{u-d1})-(\ref{u-d}) one can transform our basic
Eq.~(\ref{asymp}) into the form
\begin{eqnarray} &&\frac{d}{d\lambda}\ln D_L(\lambda)+\frac{2\lambda}{1-\lambda^{2}}L=\nonumber\\ &=&\frac{i}{\pi(1-\lambda^2)} \int_{\Xi}\mathrm{tr}\, \left[
 \Theta^{-1}(z) \frac{d}{dz}\Theta(z)\sigma_3 \right]dz, \label{asymp30}
\end{eqnarray}
here $\lambda \neq \pm 1,\, \, \pm \lambda_{m}$.
Using the same arguments as for Eq.~(\ref{detTheta}), one
can see that
$$\mathrm{tr}\, \left[\Theta^{-1}(z)  \frac{d}{dz} \Theta(z)\sigma_3 \right]=$$
\begin{equation}\label {tracetheta}= \frac{1}{cw(z)}\frac{d}{d\beta}\ln 
\left[ \theta_3\left(\beta(\lambda)+\frac{\sigma \tau}{2}\right)\theta_3\left(\beta(\lambda)-\frac{\sigma
\tau}{2}\right)\right].\end{equation}
This relation allows further simplification of Eq.~(\ref{asymp}).
Indeed, we have,
\begin{eqnarray}
&&\frac{d}{d\lambda}\ln D_L(\lambda)
+\frac{2\lambda}{1-\lambda^{2}}L=\nonumber\\
&=& \frac{d}{d\lambda}
\ln \left[ \theta_3\left(
\beta(\lambda)+\frac{\sigma \tau}{2}\right)
\theta_3\left(\beta(\lambda)-\frac{\sigma \tau}{2}\right)
\right], \label{asymp31}
\end{eqnarray}
here $\lambda \neq \pm 1,\, \, \pm \lambda_{m}$.
Taking into account the
fact that as $\lambda \to \infty$, $D_L(\lambda)\to
(-1)^{L}\lambda^{2L}$, we obtain from Eq.~(\ref{asymp31})
the following asymptotic representation for the
{\it Toeplitz determinant} $D_L(\lambda)$:
$$D_L(\lambda) =\frac{(-1)^{L}}{\theta^{2}_{3}\left(\frac{\sigma\tau}{2}\right)}
(\lambda^2 - 1)^{L} \theta_{3}\left(\beta(\lambda) +
\frac{\sigma\tau}{2}\right) \theta_{3}\left(\beta(\lambda) -
\frac{\sigma\tau}{2}\right)$$
here $\lambda$ lies outside of fixed but arbitrary neighborhoods
of the points $\pm 1$ and $\pm \lambda_{m}$, $ m \geq 0$.

It is worth noticing that the  asymptotic representation for the
   Toeplitz determinant above shows that, in the large $L$ limit,
   the points $\lambda_{m}$ (\ref{zeros}) are double zeros  of the
   $D_L  (\lambda) $.
This  suggests that in the
large $L$ limit the eigenvalues $\nu_{2m}$ and  $\nu_{2m+1}$ from 
(\ref{eaap1}), (\ref{vd})  merge:
$\nu_{2m}, \nu_{2m+1} \to \lambda_{m}.$
In turn it  indicates the degeneracy of the spectrum of the
matrix ${\bf B}_{L}$   and  an appearance
of an {\bf extra symmetry} in the large $L$ limit.

Substituting Eq.~(\ref{asymp31}) into the original equation Eq.~(\ref{eaa}),
and deforming the original contour of integration to the contour
$\Gamma$ as indicated  in Fig.~\ref{fig1} we arrive at the following
expression for the {\it entropy}:
\begin{equation}\label{33}
S(\rho_{A})=
\end{equation}
$$
= \frac{1}{2}\int_{1}^{\infty}\ln
\left(\frac{\theta_{3}\left(\beta(\lambda) + \frac{\sigma \tau}{2}\right)
\theta_{3}\left(\beta(\lambda) - \frac{\sigma \tau}{2}\right)}
{\theta^{2}_{3}\left(\frac{\sigma \tau}{2}\right)}\right)\, d\lambda
$$
This is a limiting expression as $L \to \infty$. We can prove that
 the corrections in
Eq.~(\ref{33})  are of order of $
O\left({\lambda_{C}^{-L}}/{\sqrt{L}}\right).$
The asymptotic expression (\ref{33}) is a theorem, we shall publish a
complete proof later.

The entropy has singularities at {\it phase transitions}. When 
$\tau \to 0$ we can use   Landen
transform (see  \cite{ww}) to get the following estimate
of the theta-function for small
$\tau$ and pure imaginary $s$:
$$\ln \frac{\theta_{3}\left(s \pm 
\frac{\sigma \tau}{2}\right)}{\theta_{3}\left(\frac{\sigma \tau}{2}\right)} = \frac{\pi}{i\tau}s^{2} \mp \pi i
\sigma s + O\left(\frac{e^{-i\pi/\tau}}{\tau^{2}} s^2\right), ~\textrm{as $\tau \to 0$}.$$
Now the leading term in the expression for  the entropy (\ref{33})
 can be replaced   by
\begin{equation}\label{4}
S(\rho_{A}) = \frac{\mathrm{i}\pi}{6\tau}+
 O\left(\frac{e^{-i\pi/\tau}}{\tau^{2}}\right)
\quad \textrm{for
$\tau \to 0$}.
\end{equation}
Let us consider two physical situations corresponding to small
$\tau$ depending on the case defined on the page 2:

\begin{enumerate}
\item{\it Critical magnetic field}:  $\gamma\neq 0$ and $h\to 2$.

This  is included  in our   Case
 $1$a  and  Case $2$, when $h> 2\sqrt{1-\gamma^2 }  $.
As $h\to 2$ the end points of the cuts  $\lambda_B \to \lambda_C$, so $\tau$ 
given by Eq.~(\ref{important}) simplifies and 
 we obtain from Eq.~(\ref{4}) that the entropy is:
\begin{eqnarray} &S(\rho_{A}) = -\frac{1}{6} \ln |2-h| + \frac{1}{3}\ln
4\gamma, \label{cardy} \\
&~~h\to 2 \qquad \gamma\neq 0 \nonumber
\end{eqnarray}
correction is $O(|2-h|\ln^{2}|2-h|) $. This limit agrees with predictions of 
conformal approach \cite{hlw,K, cardy}.
 The
 first term in the right hand side of (\ref{cardy})
can be represented as $(1/6)\ln \xi$, this confirms a conjecture of
 \cite{cardy}.  The correlation length $\xi$
  was  evaluated in \cite{barouch, mccoy}.

\item{\it An approach to $XX$ model}:  $\gamma\to 0$ and $h<2$: It
is included in Case $1$b, when  $0<h<2\sqrt{1-\gamma^2} $. Now
 $\lambda_B \to \lambda_C$ and  $\lambda_A \to \lambda_D$, we can calculate $\tau$ explicitly. The entropy becomes:
\begin{eqnarray}
&S^{0}(\rho_{A}) = -\frac{1}{3} \ln \gamma + \frac{1}{6}{\ln
(4-h^2)}+\frac{1}{3}\ln 2, \nonumber\\
&\gamma\to 0 \qquad h<2
\end{eqnarray}
correction is $O(\gamma\ln^2 \gamma) $. This agrees with \cite{jin}.
\end{enumerate}

It is interesting to compare this critical behavior to Lipkin-Meshkov-Glick 
model. It is similar to $XY$ model but each pair of spins interact with equal 
force, one can say that it is a model on a complete graph. The critical behavior
in  Lipkin-Meshkov-Glick was described in \cite{julien}, it is similar to $XY$,
but actual critical exponents are different.

{\it Note.} We can integrate over the original contour $\Gamma'$ of
 Fig.~\ref{fig1}
after substituting Eq.~(\ref{asymp31}) into the  equation Eq.~(\ref{eaa}).
This will give  an  alternative representation
for the entropy $S(\rho_{A})$ in terms of an infinite series:
\begin{equation}\label{3333}
S(\rho_A) = 2\sum_{m=0}^{\infty} H(\lambda_m) =\sum_{m=-\infty}^{\infty}
(1+\lambda_{m})\ln \frac{2}{1+\lambda_{m}},
\end{equation}
here the numbers $\lambda_{m}$ are defined in
Eq.~(\ref{zeros}). This representation is similar to  Eq.~(\ref{eaap1}).

{\it Remark}. These numbers $\lambda_{m}$  satisfy an estimate:
$$
|\lambda_{m+1} - \lambda_{m}| \leq 4\pi \tau_{0} \quad \mbox{with} \quad \tau_{0} = -i\tau.
$$
This means that  $(\lambda_{m+1} - \lambda_{m}) \to 0$ as $\tau \to 0$
for every $m$. This is useful for understanding of  large $L$ limit of 
the $XX$ case
corresponding  to $\gamma \to 0$, as  considered in \cite{jin}.  The estimate 
explains why in the $XX$ case  the singularities of the
logarithmic derivative of the Toeplitz determinant 
 $d\ln D_L(\lambda)/ d\lambda $ form
a cut along the interval $[-1, 1]$,
while in the $XY$ case it has  a discrete set of poles at points $\pm \lambda_{m}$ of Eq.~(\ref{zeros}). 

It would be interesting to generalize our approach to the a new class
of quantum spin chains introduced recently by J.  Keating and F. Mezzadri, 
while study matrix models
 \cite{keat}.

\section{Summary}

Our main result is the theorem that  the  expression for the 
limit (as $L\to \infty$) of the  entropy of a block of  $L$  neighboring spins on the infinite lattice 
is given by  formula (\ref{33}).  We are preparing a 
larger file with all the details of the  proof.
We can  change variables and represent it in the form:
\begin{equation}
S(\rho_{A})= \label{333}
\end{equation}
$$
= \frac{\pi}{2} \int_{0}^{\infty}\ln
\left(\frac{\theta_{3}\left(\mathrm{i} x + \frac{\sigma \tau}{2}\right)
\theta_{3}\left(\mathrm{i} x -\frac{\sigma \tau}{2}\right)}
{\theta^2_{3}\left(\frac{\sigma \tau}{2}\right)}\right)
\frac{dx}{\sinh^2(\pi x)} 
$$
We remind that  $\sigma=0$ for Case 2 and $\sigma=1$ in Case 1, see page 2.
$S(\rho_{A})$ implicitly depends on $\gamma $ and $h$
introduced in (\ref{xxh}) by means of $\tau$ defined in Eq.~(\ref{important}),
this
$\tau$ is  a ratio of periods for the theta function
$\theta_3(z)$, see (\ref{jac}). The $\gamma$ and $h$ define $\lambda_A$,  $\lambda_B$,  
$\lambda_C$ and  $\lambda_D$ by means of  Eqs.~(\ref{ldef1}), (\ref{ldef2}),
Fig.~{\ref{fig2}} and its caption. Finally  Eq.~(\ref{important}) and Eqs.~(\ref{33})
or (\ref{333}) define $S(\rho_{A})$  .

\section{Appendix}
After our paper appeared in quant-ph,  I.Peschel \cite{pes} simplified our 
expression for the entropy in the Cases 1a and 2.
He used the approach of \cite{cardy}.
He showed that in these cases our formula (\ref{3333}) 
is equivalent to formula (4.33) of  \cite{cardy}. Moreover,
I. Peschel was able to sum it up into the following expressions for the entropy. 
\begin{eqnarray}
 S=   \frac {1} {6} \left [\;\ln{ \left (\frac {k^2} {16 k'}\right )} + \left (1-\frac {k^2} {2}\right )
         \frac {4 I(k) I(k')} {\pi} \right ] + \ln\;2 , \nonumber
    \end{eqnarray}
in {\bf Case 1a}, and
\begin{eqnarray}
   S=  \frac {1} {12} \left [\;\ln{ \frac {16} {(k^2 k'^2)}} + (k^2-k'^2)
         \frac {4 I(k) I(k')} {\pi} \right ],
   \end{eqnarray}
in {\bf Case 2 }. Here, $I(k)$ denotes the complete elliptic integral of 
the first kind, $k'=\sqrt{1-k^2}$, and 
\begin{eqnarray}
 k= \left \{ \begin {array} {c} \sqrt{(h/2)^2+\gamma^2-1}\; /\; \gamma , \;\;\;\mbox{Case 1a} \\ [0.3cm]
       \gamma\; / \;\sqrt{(h/2)^2+\gamma^2-1} ,\;\;\; \mbox{Case 2}  \end{array} \right.
  \label{mod}
\end{eqnarray}

In our work, we have shown, in particular, that equation 
(\ref{3333}) is valied in Case 1b as well. 
Therefore, we can apply the summation
procedure of \cite{pes} and obtain that  in {\bf Case 1b}
\begin{eqnarray}
 S=   \frac {1} {6} \left [\;\ln{ \left (\frac {k^2} {16 k'}\right )} + \left (1-\frac {k^2} {2}\right )
         \frac {4 I(k) I(k')} {\pi} \right ] + \ln\;2 \nonumber
 \end{eqnarray}
with $k'=\sqrt{1-k^2}$, and 
\begin{eqnarray}\label{mod1b}
k=\sqrt{\frac{1-(h/2)^2-\gamma^2}{1-(h/2)^2}}\;
\end{eqnarray}

{\it Acknowledgments.} We  thank  B. McCoy, 
P. Deift, P. Calabrese and I. Peschel
for useful  discussions. This work was supported by NSF Grants
DMR-0302758, DMS-0099812 and DMS-0401009. The first co-author  
thanks B. Conrey, F. Mezzardi, P. Sarnak, and N. Snaith - the organizers of the
2004 program at the Isaac Newton Institute
for Mathematical Sciences on Random Matrices, where part of this work
was done, for an extremely stimulating research environment
and hospitality during his visit.

\end{document}